\documentclass[reprint,aps,prl,amsmath,amssymb,showpacs]{revtex4-1}

\usepackage[english]{babel}
\usepackage{siunitx}
\usepackage{graphicx}
\usepackage{xcolor}
\definecolor{linkcolor}{rgb}{0,0,0.6}
\usepackage[colorlinks=true, linkcolor= linkcolor, citecolor= linkcolor, urlcolor= linkcolor]{hyperref}

\begin{document}


\title{Brownian granular flows down heaps}


\author{Antoine B\'{e}rut}
 \email{antoine.berut@univ-lyon1.fr}
\altaffiliation{Present address: Univ Lyon, Universit\'e Claude Bernard Lyon 1, CNRS, Institut Lumi\`ere Mati\`ere, F-69622 Villeurbanne, France}
\affiliation{Aix Marseille Univ, CNRS, IUSTI, Marseille, France\\}

\author{Olivier Pouliquen}
\email[]{olivier.pouliquen@univ-amu.fr}
\affiliation{Aix Marseille Univ, CNRS, IUSTI, Marseille, France\\}

\author{Yo\"{e}l Forterre}
\email[]{yoel.forterre@univ-amu.fr}
\affiliation{Aix Marseille Univ, CNRS, IUSTI, Marseille, France\\}


\date{\today}

\begin{abstract}
We study the avalanche dynamics of a pile of micrometer-sized silica grains in water-filled microfluidic drums. Contrary to what is expected for classical granular materials, avalanches do not stop at a finite angle of repose. After a first rapid phase during which the angle of the pile relaxes to an angle $\theta_c$, a creep regime is observed where the pile slowly flows until the free surface reaches the horizontal. This relaxation is logarithmic in time and strongly depends on the ratio between the weight of the grains and the thermal agitation (gravitational P\'eclet number). We propose a simple one-dimensional model based on Kramer's escape rate to describe these Brownian granular avalanches, which reproduces the main observations.
\end{abstract}




\maketitle


Granular flows usually refer to the flows of macroscopic solid particles, typically larger than \SI{100}{\micro\meter}  \cite{Andreotti2013}. As such, they are often  considered as an archetype of athermal-disordered medium, for which thermal agitation is negligible \cite{liu2010jamming,bi2015statistical}. In these systems, the only source of particle fluctuation is the flow itself. As a result, there exists a flow threshold below which the medium can sustain a finite shear stress without flowing.  This is why a pile of grains must be inclined above a critical angle in order to flow, a property of major consequence for geophysical flows and industrial applications.

The rheology of athermal (non-Brownian) granular flows and dense suspensions has been studied in great detail over the last decades \cite{forterre2008flows,denn2014rheology,guazzelli2018rheology}.  By contrast, little is known about the behavior of granular flows when the particles become small enough that thermal fluctuations are no longer negligible. The role of thermal fluctuation has been extensively studied in hard colloidal suspensions made of very small particles immersed in a liquid~\cite{Mewis2012,ikeda2013disentangling}. However, these studies mainly concern rheology under volume-imposed conditions where the control parameter is the volume fraction of the suspension, while the flow of granular materials under gravity occurs under pressure-imposed conditions. Two different types of approaches can be found in the literature to study agitated grains under the influence of gravity.  On the one hand, for macroscopic granular materials, the application of external vibrations has been  used as a temperature analogue to study the effect of particle agitation on compaction \cite{Knight1995,philippe2002compaction} or  flow behavior \cite{Sanchez2007,dijksman2011jamming,Gaudel2016}. However, the injection of mechanical energy in a dissipative system such as a granular material is a highly out-of-equilibrium process, making  the thermal analogy non-trivial \cite{aumaitre2004energy,visco2006fluctuations}. On the other hand, the competition between gravity and thermal agitation has been studied in heavy colloids, but most often in the regime where Brownian fluctuations are dominant  compared to gravitational forces ~\cite{Royall2005,Piazza2014,Dullens2006,Thorneywork2017}. Only few numerical works have considered the opposite dense regime where agitation is small compared to particle pressure, and how it is related to the classical athermal regime of granular media~\cite{wang2015constant,Trulsson2015}.  

In this Letter, we study the avalanche dynamics of a granular medium made of micrometer-sized particles, which are heavy enough to settle in the surrounding fluid and form a pile, but also small enough to be sensitive to thermal agitation. We find that, unlike a macroscopic granular material, the avalanche of such a Brownian granular material does not stop at a finite pile angle, but slowly creeps until its free surface becomes horizontal~--~a result we rationalize within a simplified model.
 
\begin{figure}[ht!]
\includegraphics[width=\columnwidth]{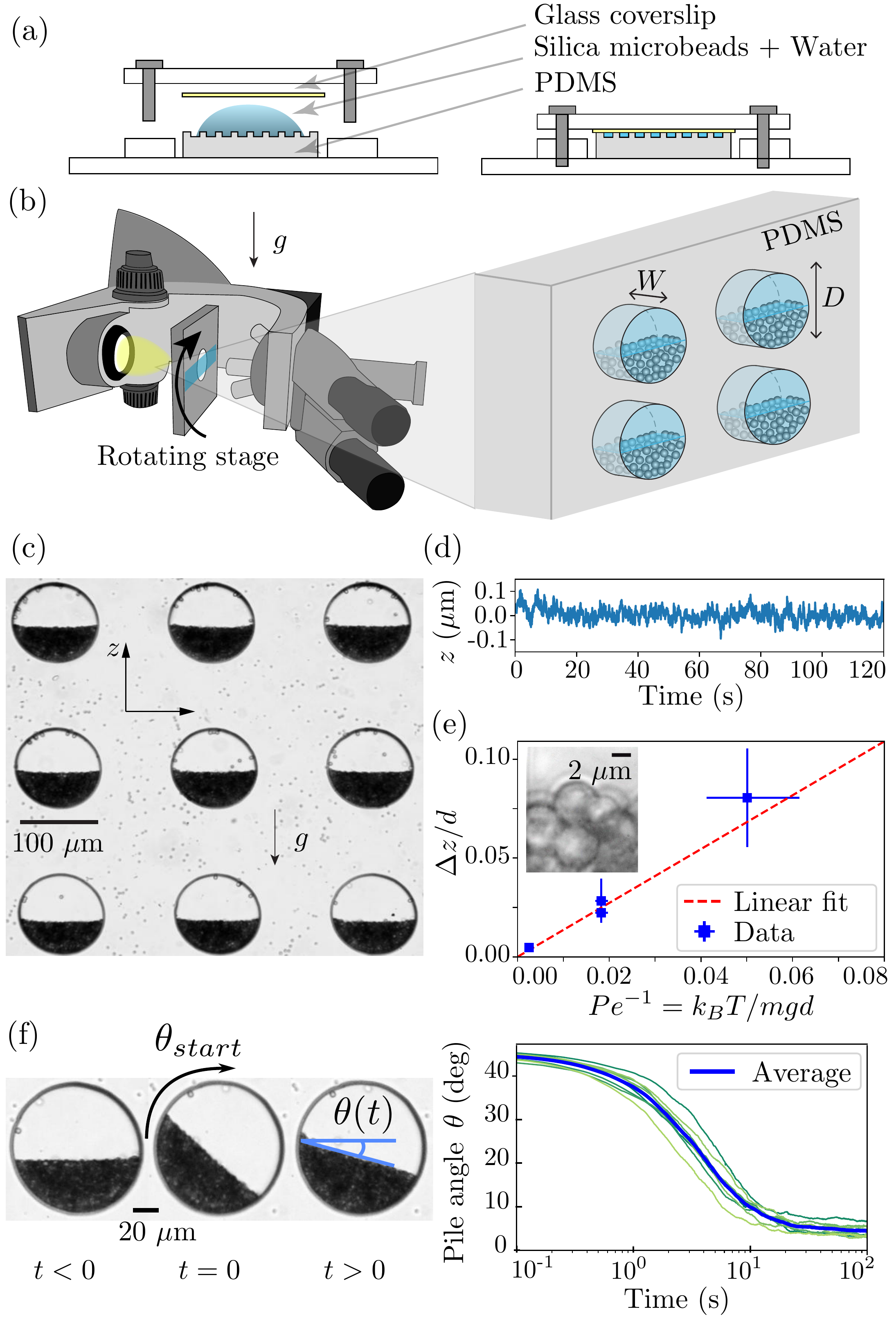}%
\caption{Brownian granular piles made of silica microparticles in water-filled microfluidic drums. (a) Sample preparation before (\textit{left}) and after (\textit{right}) sealing. (b) Sketch of the experimental set-up with inclined microscope and rotating stage to trigger avalanches ($D=\SI{100}{\micro\meter}$, $W=\SI{50}{\micro\meter}$). (c) Wide view of the water-filled PDMS drums half-filled with \SI{4.4}{\micro\meter} silica particles under gravity.  (d) Vertical thermal fluctuations of a single silica particle (diameter $d= \SI{4.4}{\micro\meter}$) at the top of the pile. (e) Amplitude of vertical fluctuations as function of the inverse gravitational P\'{e}clet number for various particle size (\textit{inset}: typical image used to track single particles at the top of the pile). (f) Avalanche generation and time evolution of the pile angle after an initial tilting $\theta_{start}=\SI{45}{\degree}$ for silica beads of diameter $d= \SI{4.4}{\micro\meter}$.  The blue curve is an average over 10 drums.  \label{fig:experimental_setup}}
\end{figure}

\textit{Experiments.} The experimental set-up is sketched in Fig. \ref{fig:experimental_setup}(a,b) and consists of thousands of microscopic cylindrical drums (diameter $D=\SI{100}{\micro\meter}$, width $W=\SI{50}{\micro\meter}$) molded in a polydimethylsiloxane (PDMS) slab (see Sup. Mat. in appendix). The drums are half-filled with inert silica microparticles dispersed in pure water and closed using a glass cover slip.  The entire device is then placed vertically on a tilted microscope, so that the observation plane contains the gravity vector. In this configuration, the particle agitation is maximal at the top of the free surface of the suspension and decreases deeper in the pile as the confining pressure induced by gravity increases. The particle agitation at the top is characterized by the gravitational P\'{e}clet number, defined by the ratio of the particle weight and the Brownian thermal forces~\cite{Russel1989}:
\begin{equation}
\mathrm{Pe} = \frac{m g d}{k_{\mathrm{B}} T}\,,
\end{equation}
where $d$ is the diameter of the particles, $m$ is their mass corrected by the buoyancy [$m = (\pi/6) d^3 \times \Delta \rho$ where $\Delta \rho \approx \SI{850}{\kilogram\per\cubic\meter}$ is the difference of density between the silica particles and the surrounding water], $g\approx \SI{9.81}{\meter\per\square\second}$ is the intensity of the gravity, $T\approx \SI{298}{\kelvin}$ is the temperature of the system and $k_{\mathrm{B}}\approx \SI{1.38e-23}{\joule\per\kelvin}$ is the Boltzmann constant. We use different batches of particles with diameter size $d$ ranging from $1.55 \pm 0.05$ to $4.40 \pm 0.24$ \si{\micro\meter}, which corresponds to P\'{e}clet numbers between 6 and 400. This range of $\mathrm{Pe}$ ensures that the particles will sediment rapidly and form a well-defined pile at the bottom of the drums. However, even the bigger ones show measurable random fluctuations induced by thermal agitation [see vertical motions on Fig.~\ref{fig:experimental_setup}(d)].  As expected, the amplitude of the vertical fluctuations $\Delta z$ of one particle at the top of the pile increases linearly with the inverse  P\'{e}clet number since  $mg\Delta z\sim k_{\mathrm{B}}T$ [Fig.~\ref{fig:experimental_setup}(e)].  In our experiments, these fluctuations are always at least one order of magnitude smaller than the particle diameter. However, we will see that this small agitation is enough to drastically modify the avalanche behavior when the drums are tilted.

Before any measurements, the sample is left in a vertical position for about $\sim$\SI{10}{\hour} to ensure that the system is well sealed. This waiting time also allows a layer of particles to stick at the bottom of the drum, forming a rough surface, which is crucial to prevent slippage of grains at the wall when the drums will be tilted. The suspension is then stirred by rapidly rotating the stage of the microscope several times, before stopping the stage for $\sim$\SI{5}{\minute} to allow the particles to settle and form a flat pile. This procedure ensures a reproducible initial state and avoids potential aging effects that could arise from long time contacts between grains~\cite{Gayvallet2002}. Finally, the pile is tilted at an initial angle $\theta_{start}$ to generate an avalanche and the temporal evolution of the pile angle $\theta(t)$ is recorded [Fig.~\ref{fig:experimental_setup}(f)].

\begin{figure}[t!]
\includegraphics[width=\columnwidth]{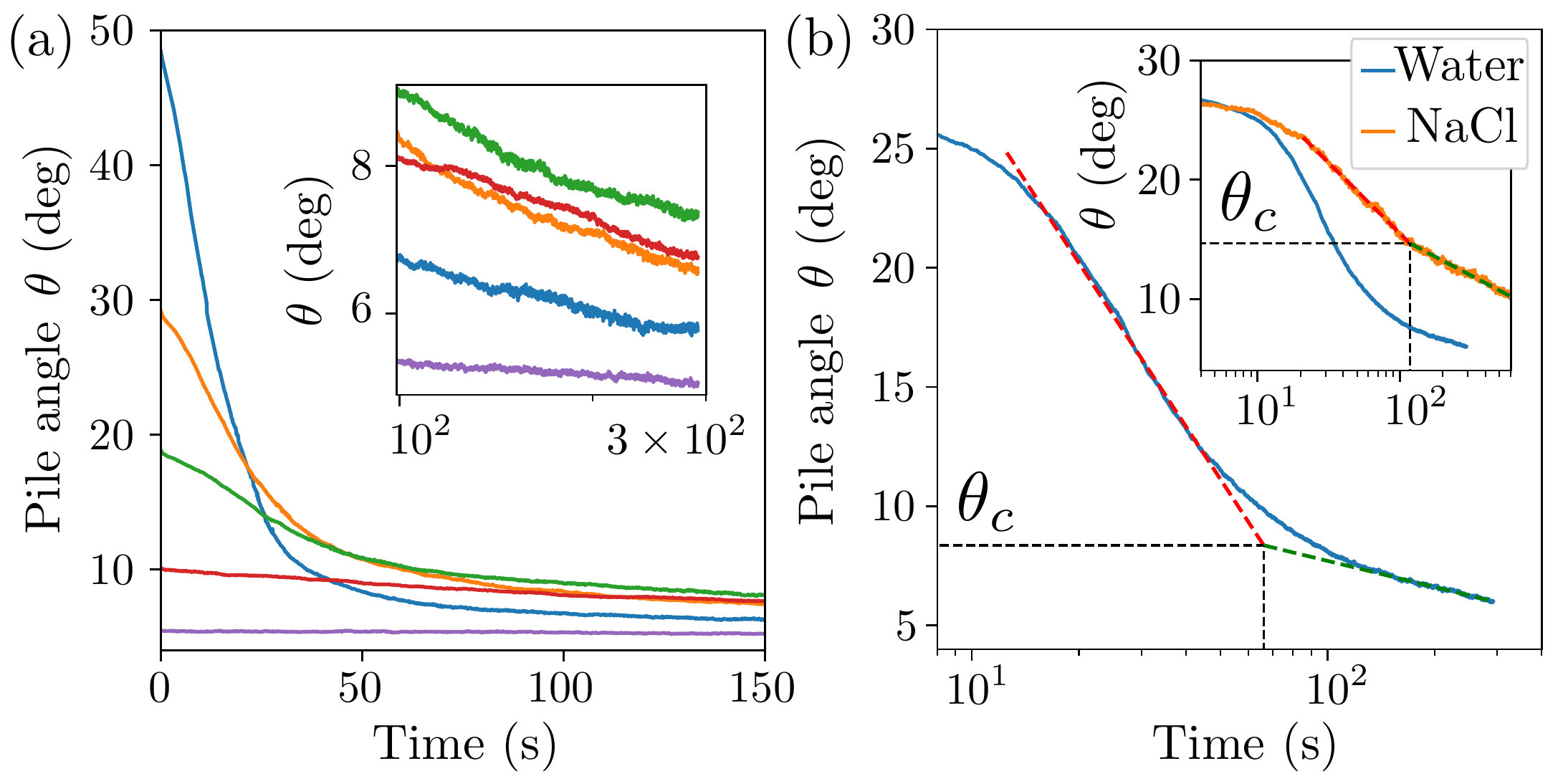}
\caption{Avalanche and creep behavior for a single particle size.  (a) Temporal evolution of the pile angle with different initial angles $\theta_{start}$ ranging from \SI{5}{\degree} to \SI{50}{\degree} for $d=\SI{2.68}{\micro\meter}$ in pure water. \textit{Inset}: Zoom on the end of the trajectories. (b) Semi-logarithm representation showing a first fast avalanching regime (red dashed line) followed by a slow creep regime (green dashed line) for $\theta_{start}=\SI{30}{\degree}$ and $d=\SI{2.36}{\micro\meter}$ in pure water. The threshold angle $\theta_c$ separating the two regimes is about \SI{8}{\degree}. \textit{Inset}: Comparison with the same particles in a NaCl ionic solution  (10$^{-2}$ mol L$^{-1}$) to screen the electrostatic repulsion between particles. In this case  $\theta_c$ is about \SI{15}{\degree}.\label{fig:avalanches_shorttimes}}
\end{figure}

First, we studied the avalanche behavior for a single particle size $d=\SI{2.68}{\micro\meter}$, corresponding to $\mathrm{Pe}\approx 55$, and for different starting angles  $\theta_{start}$ ranging from \SI{5}{\degree} to \SI{50}{\degree} [Fig.~\ref{fig:avalanches_shorttimes}(a)]. After a short avalanche phase during which the angle of the pile angle quickly relaxes, the main observation is that the medium seems to continue to flow over a long period, even at very small angles [inset of Fig.~\ref{fig:avalanches_shorttimes}(a)]. This is in striking contrast with the avalanche behavior of an immersed macroscopic granular material, for which the medium stops at a well-defined angle of repose~\cite{CourrechduPont2003,Rondon2011}. Interestingly, a flow is still observed when $\theta_{start}=\SI{5}{\degree}$, i.e. when the initial angle is well below the typical angle of repose for macroscopic granular materials (typically between 20 to \SI{30}{\degree})~\cite{Andreotti2013}, and even below the angle of repose found for frictionless hard spheres  ($\sim 6\si{\degree}$)~\cite{Peyneau2008,Clavaud2017}. This observation is reproducible  and was observed for all the particle sizes considered in this study~\cite{OpenData}. It also seems rather insensitive to aging, as \SI{48}{\hour} old samples exhibit nearly the same avalanche dynamics as the one observed at their first use.

An examination of the avalanche dynamics using a semilogarithmic plot confirms that the flow is composed of two very different regimes: a fast avalanche regime for $\theta>\theta_c$ and a slow creep regime for $\theta<\theta_c$ [Fig.~\ref{fig:avalanches_shorttimes}(b) for $d=\SI{2.36}{\micro\meter}$ and $\theta_{start}=\SI{30}{\degree}$]. The threshold angle $\theta_c$ is reminiscent of the angle of repose of a macroscopic granular material.  For our silica particles, $\theta_c \approx \SI{8}{\degree}$, and does not seem to vary with the size of the particles.  Such a low value of the angle of repose indicates that the silica particles interact through almost frictionless contacts in pure water. This is due to the presence of a short-range repulsion force of electrostatic origin between the negatively charged surfaces of the particles, which is large enough to sustain their weight even for the particles at the bottom of the drum where the pressure is maximal~\cite{Clavaud2017}. This interpretation  is consistent with the fact that $\theta_c$ can be increased to $\sim$\SI{15}{\degree} if the particles are immersed in a solution of NaCl at 10$^{-2}$ mol L$^{-1}$  [orange curve, inset of Fig.~\ref{fig:avalanches_shorttimes}(b)]. In this case, the repulsive force is screened by the ions of the solution~\cite{Israelachvili2011} and the contacts between particles become frictional, thereby increasing the value of the angle of repose~\cite{Clavaud2017}. Interestingly, the phenomenology observed with the frictionless particles in pure water remains when  particles are frictional. Below the critical angle $\theta_c$, the medium does not stop like a classical granular material but slowly creeps. 
This creep likely comes from the thermal agitation of the particles, which enables particles to jump and move with respect to their neighbors.

\begin{figure}[ht!]
\includegraphics[width=\columnwidth]{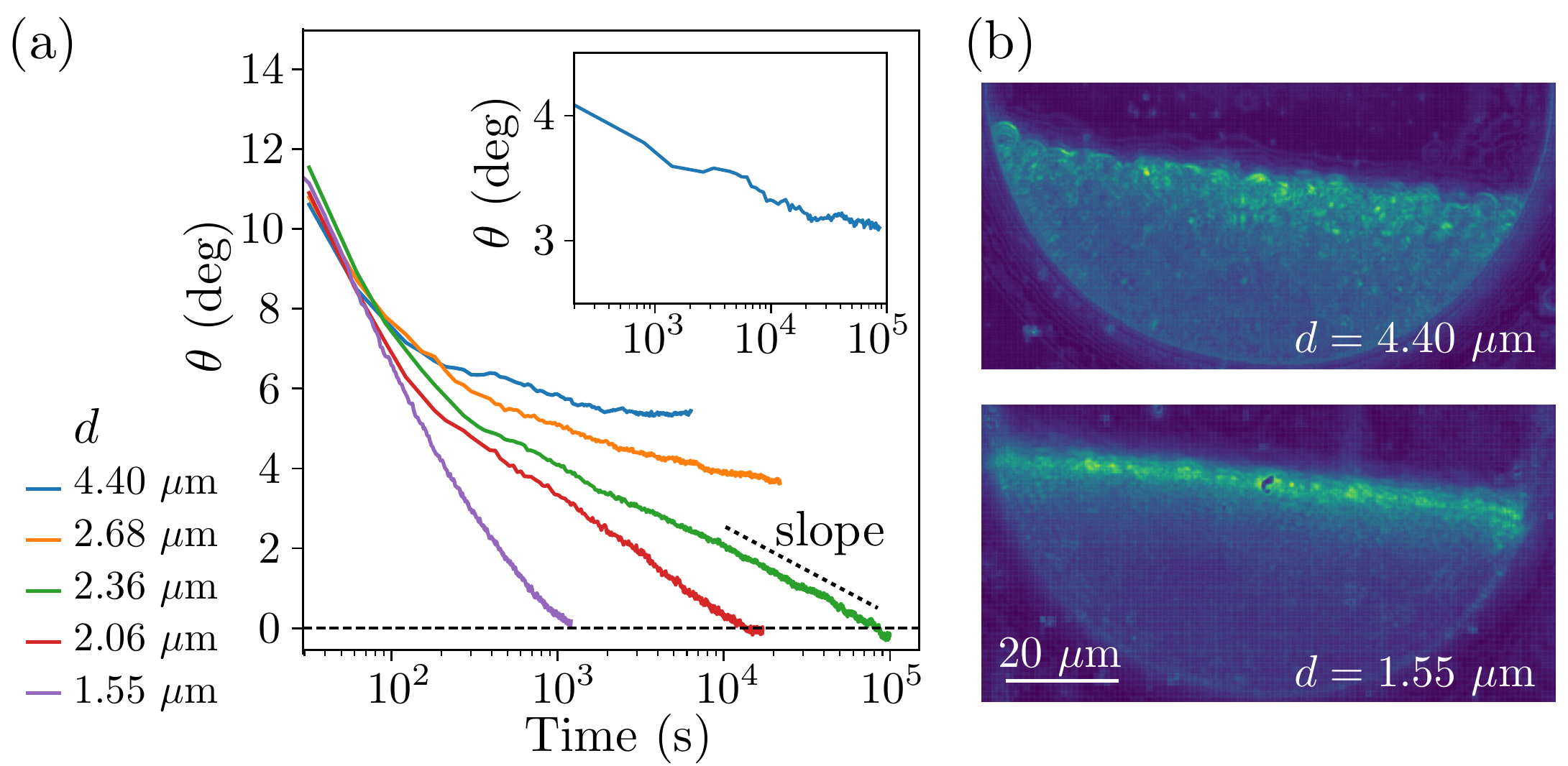}
\caption{Thermal creep regime. (a) Long-time temporal evolution of the pile angle for various particle sizes corresponding to $\mathrm{Pe}=397$ ($d=\SI{4.40}{\micro\meter}$), $\mathrm{Pe}=55$ ($d=\SI{2.68}{\micro\meter}$), $\mathrm{Pe}=33$ ($d=\SI{2.36}{\micro\meter}$), $\mathrm{Pe}=19$ ($d=\SI{2.06}{\micro\meter}$), and $\mathrm{Pe}=6$ ($d=\SI{1.55}{\micro\meter}$) ($\theta_{start} = \SI{15}{\degree}$). \textit{Inset}: Long-time behavior for \SI{4.40}{\micro\meter} particles, with lower initial angle. (b) Heat map of particles displacement in the creep regime for two particle sizes. The map is obtained by computing image differences with a small time step (\SI{1}{\second} for \SI{1.55}{\micro\meter} particles and \SI{2}{\second} for \SI{4.40}{\micro\meter} particles) and averaging over \SI{1}{\minute}. The light colors correspond to area where the particles are moving.\label{fig:avalanches_longtimes}}
\end{figure}

To confirm this picture, we studied the effect of the P\'{e}clet number on the long-time avalanche behavior using different particle sizes, starting at the same initial inclination angle $\theta_{start}=\SI{15}{\degree}$ [Fig.~\ref{fig:avalanches_longtimes}(a)]. The fast avalanche regime above $\theta_c$ seems rather independent of $\mathrm{Pe}$ as the different curves collapse pretty well on short times. On the contrary, the dynamics of the creep regime slows dramatically as the P\'{e}clet increases, i.e. when the particle weight increases compared to the particle agitation.  Below $\theta_c$, we observe a relaxation of $\theta$ that is logarithmic in time on several decades. For low $\mathrm{Pe}$ (large relative agitation), the free surface of the pile creeps rapidly,  until it abruptly stops when it reaches the horizontal. When $\mathrm{Pe}$ increases, the slope of the logarithmic regime decreases, meaning that the time needed to reach $\theta = \SI{0}{\degree}$ becomes longer and longer. In practice, for the largest particle sizes, this asymptotic horizontal state could not be observed, as evaporation from the side of the PDMS limits the duration of the experiments to less than $\sim$ 48h. In the limit $\mathrm{Pe} \gg 1$, one expects to recover the macroscopic granular behavior with no avalanche motion below the angle of repose. Here, creep below the angle of repose was still observed for the highest $\mathrm{Pe}$ studied [inset of Fig.~\ref{fig:avalanches_longtimes}(a)].  

\begin{figure}[t!]
\includegraphics[width=\columnwidth]{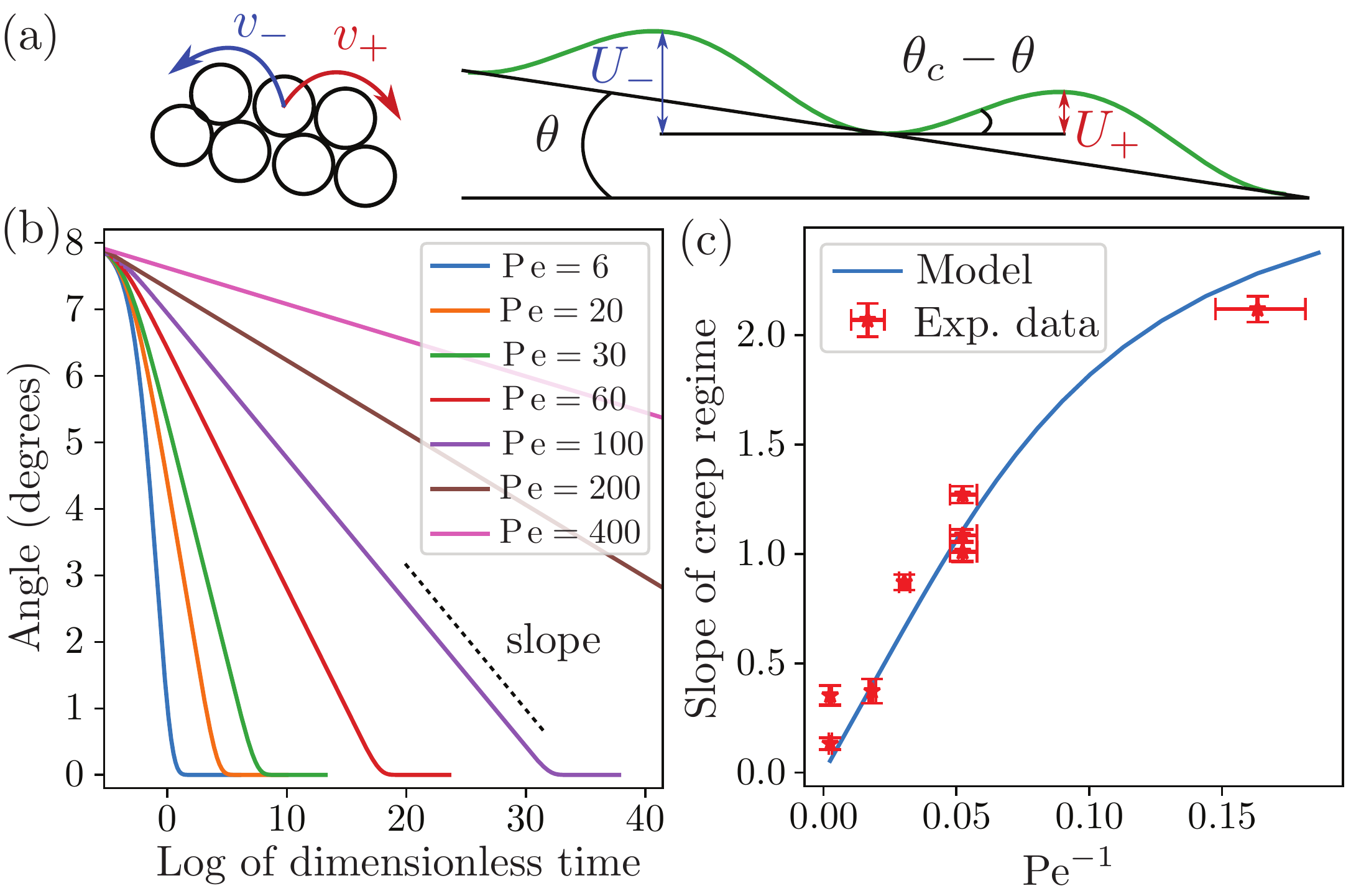}%
\caption{Thermally-activated avalanche model. (a) Sketch of the potential felt by a particle at the top of the inclined pile. (b) Avalanches dynamics in the creep regimes predicted by the model for various values of $\mathrm{Pe}$, as a function of $\log(t/\tau)$ ($\alpha = 2.64$).  (c) Comparison of the logarithmic regime slopes between the experimental data and the model with $\alpha = 2.64$. Each experimental data corresponds to a sample. Horizontal errorbars comes from the dispersion of particle's diameters. Vertical errorbars are given by the dispersion of the slopes measured in the different drums of each sample.\label{fig:model}}
\end{figure}

\textit{Model.} To describe the previous phenomenology, we propose a simple avalanche model for a Brownian granular material in the creep regime. The model is based on the assumption that a particle at the top of the pile can be described as a single Brownian particle in a frozen energy landscape imposed by the underlying particles, as sketched in Fig.~\ref{fig:model}(a). This single particle approach that neglects collective effects is motivated by the fact that flow in the creep regime is limited to a few layer of grains at the top of the pile, typically one or two particle sizes only  [Fig.~\ref{fig:avalanches_longtimes}(b)]. Considering a particle going down the pile, we expect that the energy barrier it faces $U_+$ is proportional to the gravitational potential diminished by the tangent of the inclination angle $\theta$. It must also vanishes for $\theta = \theta_c$ because the particles are free to fall without thermal motion if the pile is inclined above the angle of repose. Similarly, the energy $U_-$ faced by a particle going up the pile is increased by $\tan \theta$. We then have $U_\pm = \alpha mgd (\tan\theta_c \mp \tan\theta)$, where $\alpha$ is a dimensionless coefficient characterizing the effective height of the energy barrier.  We next assume that the particle can cross barriers of potential energy $U$ with a probability $p$ that follows Kramer's theory: $p\propto \exp(-U/k_\mathrm{B}T)$~\cite{Kramers1940,Hanggi1990}. The average velocity $v_+$ (resp. $v_-$) of a particle going down (resp. up) the pile is then given by:
\begin{equation}
v_\pm = f d \exp \left[ -\alpha \mathrm{Pe} (\tan\theta_c \mp \tan\theta) \right],
\end{equation}
where $f$ is a rate prefactor, which is dimensionally proportional to $\Delta \rho g d/\eta$, with $\eta$ the viscosity of the fluid surrounding the particle.
From volume conservation, the rate of change of the pile angle is related to the avalanche flow rate by: $1/2 \times (D/2)^2 \mathrm{d}\theta/\mathrm{d}t = h(v_+ - v_-)$, where $h \sim d$ is the height of the layer of flowing particles.  We then obtain the following law for the evolution of the pile angle:
\begin{equation}
\frac{\mathrm{d}\theta}{\mathrm{d}t} = -\frac{1}{\tau} \operatorname{e}^{-\alpha \mathrm{Pe} \tan \theta_c} \sinh \left(\alpha \mathrm{Pe} \tan \theta \right),
\end{equation}
where the timescale $\tau \propto D^2\eta/(\Delta\rho g d^3)$ depends on the particle's weight, the viscosity of the fluid and the dimension of the drum. Since pile inclinations in the creep regime are small, we assume that $\tan \theta \approx \theta$ and solve the equation with the initial condition $\theta\vert_{t=0} = \theta_c$, which gives:
\begin{equation}
\theta(t) = \frac{2}{\alpha\mathrm{Pe}} \, \mathrm{arcoth} \left[ \exp \left( \frac{t}{\tau} \alpha \mathrm{Pe} \, \operatorname{e}^{- \alpha\mathrm{Pe} \theta_c} \right) \,\coth (\alpha \mathrm{Pe} \theta_c /2) \right].
\end{equation}

In Fig.~\ref{fig:model}(b), this solution is plotted as a function of $\log(t/\tau)$ for different values of $\mathrm{Pe}$. The model reproduces the  logarithmic dynamics  on long time of the pile angle relaxation, with a slope that decreases when $\mathrm{Pe}$ increases. It also recovers the sharp change of regime when $\theta$ approaches \SI{0}{\degree}.  In the limit of large $\mathrm{Pe}$, the solution can be simplified to:
\begin{equation}
\theta(t) \approx \theta_c - \frac{1}{\alpha\mathrm{Pe}} \log \left( \alpha \mathrm{Pe} \frac{t}{2\tau} \right)\,,
\end{equation}
provided that $\theta$ is not too close from $\theta_c$ or \SI{0}{\degree}. The slope of the logarithmic regime thus evolves as $1/\mathrm{Pe}$ and the time $t_{\rm stop}$ to relax to the horizontal is given by an Arrhenius law $t_{\rm stop}=(2\tau/\alpha \mathrm{Pe})\exp(\alpha \theta_c\mathrm{Pe})$.

Fig.~\ref{fig:model}(c) compares the slopes of the logarithmic regime predicted by the model  with the slopes measured experimentally in the creep regime. The best agreement is obtained using $\alpha = 2.64$, but the agreement remains reasonable for $2 < \alpha < 3$.  In the framework of the model, such values of $\alpha>1$ mean that the particle needs to overcome a geometrical barrier higher than the diameter of its neighbors. This large value of $\alpha$ shows the limit of our over-simplistic one-dimensional approach, which does not take into account collective or 3D geometrical effects. 

Once $\alpha$ is fixed, the values of $\tau$ can be estimated by matching the time $t_{\rm stop}$ needed for the pile to reach $\theta = \SI{0}{\degree}$ in the model [Fig.~\ref{fig:model}(b)] and in the experiments [Fig.~\ref{fig:avalanches_longtimes}(a)]. We made the comparison for the three smaller sizes (1.55, 2.06 and \SI{2.36}{\micro\meter}) only, as for larger particles the creep is very slow and $t_{\rm stop}$ can only be obtained by extrapolation,  resulting in large uncertainties.  Experimentally, we find  $\SI{30}{\second} < \tau < \SI{600}{\second} $, in reasonable agreement with the scaling $\tau \propto D^2\eta/(\Delta\rho g d^3)$ that gives values between \SI{90}{\second} and  \SI{320}{\second}  for those particle sizes. Our model therefore reproduces the basic features of the Brownian creeping avalanches regime  observed experimentally, even though a more complex one would be require to fully describe the flow dynamics.

\textit{Conclusion.} 
In this letter, we have studied the avalanche dynamics of a ``Brownian granular material'', made of nearly frictionless grains small enough to be sensitive to thermal agitation. This granular material exhibits a slow creep regime below its angle of repose, that does not exist in its athermal counterpart. In this regime, the pile angle shows a logarithmic relaxation to zero that dramatically depends on the P\'{e}clet number. This phenomenon is reminiscent of the relaxation observed in macroscopic granular materials submitted to external perturbations, such as vibration~\cite{Sanchez2007,dijksman2011jamming}, thermal cycling~\cite{Divoux2008}, shear below yield stress~\cite{Nguyen2011}, or fluid-injection below the fluidization threshold~\cite{houssais2019soil}. It is also analogous to the mechanism observed in the gravity sensing cells of plants, where the gravisensors made of tiny heavy grains are agitated by the cytoskeleton activity to promote their mobility under gravity~\cite{berut2018gravisensors}. To describe this creep regime,  we have proposed a one-dimensional thermally-activated model based on a Kramer 's escape rate. Despite its simplicity, the model gives good agreement with experimental data with only one free-parameter.

Our results are a first step toward a better understanding of the flow of Brownian granular material under imposed pressure $P$. Experimentally, a flow geometry better controlled than drum flow, such as the inclined plane, could be used to vary the two main rheological parameters in this case, namely the stress P\'eclet number $\mathrm{Pe}=Pd^3/k_BT$ and the viscous number $J=\eta\dot{\gamma}/P$~\cite{wang2015constant,Trulsson2015}. Making the link with the rheology of dense Brownian suspensions usually studied at constant volume would be interesting as well.  For Brownian hard spheres, a glass transition is predicted at a packing fraction lower than the critical jamming packing fraction of athermal systems~\cite{ikeda2013disentangling}.  Whether this glass transition observed at constant volume has counterpart when the medium flows at constant pressure  is an open question, which we are currently investigating. \bigskip

\begin{acknowledgments}
This work was supported by the European Research Council (ERC) under the European Union's Horizon 2020 research and innovation programme (grant agreement N$^\circ$647384) and by  the French National Agency (ANR) under the program Blanc Grap2 (ANR-13-BSV5-0005-01), Labex MEC (ANR-10-LABX-0092) and A*MIDEX project (ANR-11-IDEX-0001-02) funded by the French government program Investissements d'avenir.
All data presented in this article are openly available in Zenodo repository: \href{https://doi.org/10.5281/zenodo.1035785}{10.5281/zenodo.1035785}
\end{acknowledgments}

\bibliography{biblio}

\newpage

\appendix
\section{Supplementary Material: Technical details about the experiment}

The matrix of cylinders is made in polydimethylsiloxane (PDMS) using standard microfluidic fabrication techniques~\cite{Duffy1998,McDonald2002,Friend2010}: a negative mold with the desired pattern was made in SU-8 photoresist at the CINaM Laboratory, then the mold was used to create a positive replica with Sylgard\textregistered 184 Silicone Elastomer Kit (standard 10:1 mixture, cured one night at \SI{60}{\celsius} in a oven). The silica particles are commercially available from Microparticles GmbH in aqueous solutions (\SI{5}{\percent}wt), and have a density $\rho \approx \SI{1850}{\kilogram\per\cubic\meter}$.

The drums are filled using the following protocol: the PDMS container is first manually cleaned with isopropyl alcohol (IPA) and rinsed with pure water (Type 1 water from Purelab\textregistered Flex dispenser). Then, it is placed in the lower part of a transparent PMMA vise and washed in pure water with a ultrasonic bath for about \SI{30}{\minute}. The dispenser is rinsed again and a few droplets of pure water are put at the top of the clean PDMS surface. Then, a small volume (between 20 and \SI{40}{\micro\liter}) of silica particle solution is injected inside the water droplet. The particles are let to sediment for $\sim$\SI{2}{minutes}. A glass coverslip, previously washed with IPA, is pressed against the PDMS thanks to the upper part of the PMMA vise that is clamped with four screws. The coverslip is pressed carefully so that the droplet spread in the container, and no air bubble form inside. Once the container is closed, it is put vertically under the microscope to find an area where no drum is leaking (i.e. where no particle is able to escape its drum to a neighboring one). 

Large scale observations are made using a microscope (Leica DM 2500P) flipped horizontally with a long working distance objectives (NPLAN EPI $10 \times/0.25$ POL) allowing to watch up to 12 drums at the same time. The sample is held on a rotational stage (M-660 PILine\textregistered) with a maximal velocity of \SI{720}{\degree\per\second} and controlled by a C-867 PILine\textregistered Controller. Images and movies are taken with a Nikon D7100.

We verified that the piles at rest show no compaction effect. The position of the interface between a pile of \SI{2.06}{\micro\meter} particles and the fluid was tracked for \SI{2}{\hour} after stirring the drums: sedimentation was observed during $\sim\SI{3}{\min}$ and no evolution was measured afterwards.

\end{document}